\documentclass[aps,prbbib,floatfix,floats]{revtex4}
\usepackage{graphicx,latexsym}
\usepackage{epsfig}
\usepackage{amsmath}
\usepackage{color}

\pdfoutput=1
\begin{document}

\title{Phase extraction in disordered isospectral shapes}
\author{Mugurel \c Tolea, Bogdan Ostahie, Marian Ni\c t\u a, Felicia \c Tolea and Alexandru Aldea}

\address{
 National Institute of Materials Physics,
POB MG-7, 77125
 Bucharest-Magurele, Romania.}

\begin{abstract}

The phase of the electronic wave function is not directly
measurable but, quite remarkably, it becomes accessible in
pairs of isospectral shapes, as recently proposed in the
experiment of Christopher R. Moon {\it et al.}, Science {\bf
319}, 782 (2008). The method is based on a special property,
called transplantation, which relates the eigenfunctions of the
isospectral pairs, and allows to extract the phase
distributions, if the amplitude distributions are known. We
numerically simulate such a phase extraction procedure in the
presence of disorder, which is introduced both as Anderson
disorder and as roughness at edges. With disorder, the transplantation can no
longer lead to a perfect fit of the wave functions, however we
show that a phase can still be extracted - defined as the phase
that minimizes the misfit. Interestingly, this extracted phase
coincides with (or differs negligibly from) the phase of the
disorder-free system, up to a certain disorder amplitude, and a
misfit of the wave functions as high as $\sim 5\%$, proving a
robustness of the phase extraction method against disorder.
However, if the disorder is increased further, the extracted
phase shows a puzzle structure, no longer correlated with the
phase of the disorder-free system. A discrete model is used,
which is the natural approach for disorder analysis. We provide
a proof that discretization preserves isospectrality and the
transplantation can be adapted to the discrete systems.

\end{abstract}

\maketitle

\section{Introduction}
Famous mathematical problems sometimes attracted a great
interest from physicists as well. One such problem was the
isospectrality debate launched by Kac in 1966 \cite{Kac} when
he asked: "Can one hear the shape of a drum"? It was known that
the spectrum uniquely determined the area and the perimeter of
a "drum", but whether it also contained the full shape
information was yet to be researched. It wasn't until 1992 that
Gordon et.al. \cite{Gordon}, in a milestone paper, answered
negatively to the famous question by finding different
(noncongruent) shapes with identical spectra. However,
isospectrality remains a high exception, only 17 such classes
of pairs being known \cite{RMP}, and it is believed that no
others exist. Soon after the paper by Gordon {\it et
al.}\cite{Gordon}, Wu {\it et al.} \cite{Wu} and Driscoll
\cite{Driscoll} found explicitly the first eigenvalues and
eigenfunctions of two - the most simple and most famous - such
isospectral shapes, called "Bilby" and "Hawk" (see Fig.1).
There has been also immediate experimental interest of
realizing such isospectral domains, by Sridar and Kudrolli
\cite{Sridar}, in a microwave cavities experiment. Other
boundary conditions for the isospectral shapes have also been
discussed in \cite{Driscoll2,Dowker,Levitin}. Mathematical and
physical aspects of isospectrality have been reviewed by Giraud
and Thas \cite{RMP} - including also pioneering contributions
of the authors.

Recently - and this was the motivation of our paper - isospectrality has found
a direct application in experimental quantum mechanics, by allowing the extraction
of the electron's phase, in a non-interferometric way.
We refer to the experiment
of Moon {\it et al.} \cite{Science}, who realized isospectral shapes
by planting CO molecules on copper surface with the use of an
STM tip.
The principle of
the phase extraction is simple: it can be shown that, if one
has two isospectral shapes, one can build the eigenfunctions of
one shape by using combination of parts from the corresponding
eigenfunction of the other shape. The procedure is called
"transplantation" (see Appendix A) and
this brings
supplementary information which are used to find the
phase distribution of the eigenfunctions.

Prior to the experiment of Moon {\it et al.} \cite{Science},
the phase measurement -in mesoscopic physics- has already
attracted a great interest. Naturally, the first experiments
used interference geometries, namely Aharonov-Bohm
interferometers with embedded quantum dots. Such experiments
(e.g. \cite{Schuster}) aimed to extract the phase of the
electron transmittance through a quantum dot and they generated
a number of intriguing questions, that are still open. We
mention briefly the universal phase lapse between resonances
(called by some authors "the longest standing puzzle in
mesoscopic physics" - e.g. \cite{Oreg,Karr}) for a
many-electron dot, or the reduced variation of the phase (with
fractions of $\pi$ on some resonances and between them) for a
few-electrons quantum dot \cite{Avinun}. As was easy to expect,
these open questions attracted many theoretical attempts to
explain them (see for instance the recent papers \cite
{Oreg,Karr,Goldstein,TNA,Rontani,Puller,Buchholz,Racec} and
references therein). The two existing phase measurement setups
(by interferometry or by use of isospectral shapes), although
different, present similarities \cite{comment1}.

In this paper, we focus on the study of isospectral shapes (in
particular the Bilby-Hawk pair) under the influence of
disorder, with an emphasis on the phase extraction procedure.
The aspect should be of interest because the
experimental conditions, for instance, are never quite perfect,
and isospectrality can only be closely approached. If, for
instance, the isospectral shapes are carefully prepared on a
flat surface, the disorder effects may come from small defects,
oscillations of the atoms due to temperature, tiny movements of
the STM measurement tip, etc. The are many ways in which
disorder or impurities can be introduced. In this paper we
present the result of averaging (the measurable quantities,
such as energy levels and wave functions amplitudes) over large
ensembles of disorder configurations of variable amplitude (diagonal Anderson disorder is considered).
With disorder, isospectrality, as well as the transplantation procedure do not hold rigourously.
One can however still define a "measurable" or "extracted" phase simulating the experimental procedure:
we will use the
(disordered averaged) wave function amplitudes and search numerically the phase distribution that leads to
the best fit after transplantation.
It will be found that this extracted phase coincides - up to negligible differences - with the phase of the "clean" shapes, if the disorder is below a given amplitude.

We consider also the effect of edge roughness, with similar
conclusions, namely that a certain degree of roughness can be
allowed. Therefore the "perfect" conditions are not necessary for a
correct phase extraction. A discrete model is used, as being the
most suitable approach for disorder analysis. The discrete approach
allows  easy tailoring of any shapes, which remain isospectral if
their continuous counterparts are isospectral (see the proof in
Appendix A). Another justification for choosing a discrete model
comes again from the phase measurement experiment of Moon et. al
\cite{Science}, where the wave functions amplitudes (used for
transplantation) were measured in a finite number of points on the
surface.

The outline of the paper is as follows: in Section II we
introduce our discrete model, Section III contains the main
results, which are summarized in Section IV. Appendix A gives a
proof that isospectrality holds in the discrete representation
and Appendix B offers a comparison between continuous and
discrete models.

\section{The discrete model}

We choose a discrete approach meaning that the isospectral shapes will be described by a
number of sites (noted with i or j) that belong to a rectangular lattice with given on-site energies $\epsilon_i$ and hopping integrals
$t_{i,j}$. A generic Hamiltonian can be written as:
\begin{equation}
H=\sum_{i}\epsilon_i|i\rangle\langle
i|+\sum_{i,j}t_{i,j}|i\rangle\langle j|.
\end{equation}
$t_{i,j}$ is chosen to be equal with 1 (or energy unit) for the nearest neighbor sites and $0$ otherwise and the diagonal energies
 $\epsilon_i$ are equal to 0 for disorder free shapes or
are given random values in the interval $[-W/2,W/2]$ in the presence of Anderson disorder with amplitude W.

 The particular shapes described by the Hamiltonian in Eq.1 are determined by the way in which the
 sites are inter-connected, two examples being the isospectral Bilby and Hawk drums shown
 in Fig.1.
The total number of sites inside each shape can be calculated if we give the value $N$ which is the number
of sites on the hypotenuse of an elementary triangle (for
Fig.1, $N=7$) \cite{sites}.

\begin{figure}[ht]
\vskip -13cm
\hskip 5cm
  \includegraphics[width=15cm]{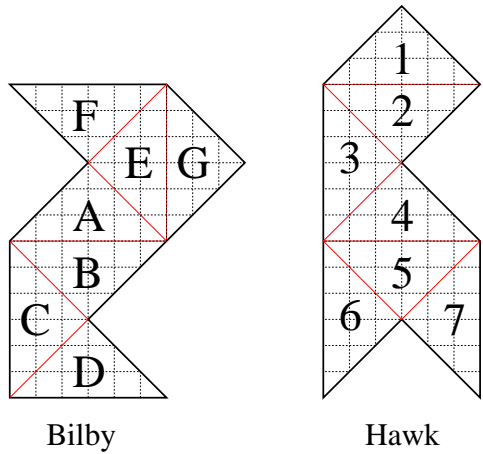}\\
  \centering
  \caption{(Color online) The isospectral shapes "Bilby" and "Hawk". They are
  represented by a discrete set of points at the intersections of the thin dashed
  lines. Each shape is divided in 7 triangles, for the purpose of transplantation
  (see description in text). The triangles are noted with $\alpha$=A,...,G for Bilby and $n$=1,...,7 for Hawk.
  One can see that some points are in the interior of triangles and
  others on the borders between triangles (thin red lines)
  or on the exterior borders (black continuous lines).
 The points on the exterior borders are considered to have infinite potential so the wave functions vanish on these points.}
\end{figure}

It is known that the continuous Bilby and Hawk shapes are isospectral.
A natural question is whether the discrete Bilby and Hawk are
also isospectral, for any discretization (i.e. any $N$), or is
the continuous limit necessary in order to achieve this property?
We prove that the first affirmation
is correct and
the isospectrality holds for any discretization. The proof is
presented in Appendix A and is based on the transplantation
method, similar to the continuous case. A further
comparison between discrete and continuous models is discussed
in Appendix B.

\begin{figure}[ht]
\vskip -5.5cm
  \includegraphics[width=8.5cm]{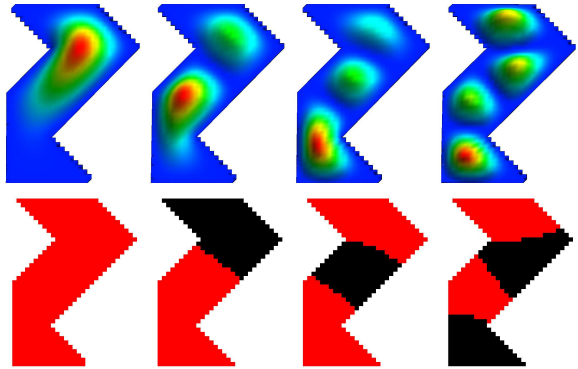}
  \centering
  \caption{(Color online) The first four eigenmodes of Bilby:
 amplitudes (first row) and phases (second row). The black and red (gray) zones have
  opposite phases (the real wave function have opposite signs).}
\end{figure}
\begin{figure}[ht]
 \vskip -1cm
  \includegraphics[width=6cm]{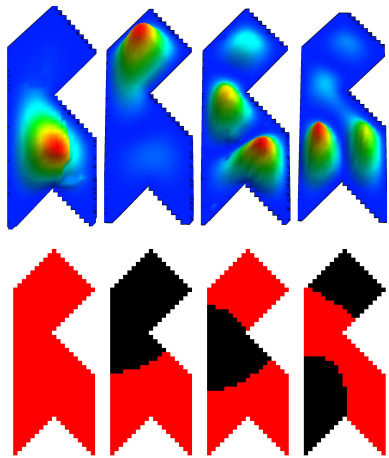}\\
  \centering
  \vskip -3cm
  \vskip 3cm
  \caption{(Color online) The same as in Fig.2, for the first four Hawk modes. }
\end{figure}

\vskip 0.5cm

We plot below the amplitudes and phases of the first four
eigenmodes of Bilby (Fig.2) and Hawk (Fig.3), corresponding to
a discretization with $N=21$. The plots are given for
completeness. While the amplitudes plots can be found in
literature for a large number of eigenmodes \cite{RMP}, we
found phase plots only in the experimental paper
\cite{Science}. A discussion is necessary regarding the plotted
phase. The shapes we discuss are closed systems and their
eigenfunctions are considered real. Therefore the phase can
only be $0$ or $\pi$ corresponding to the sign $+$ or $-$ of
the wave function on a particular site (see also the discussion
in \cite{Science}). Obviously, the phase of a wave function is
defined up to a constant, so what is relevant is the relative
phase difference between the points on the surface. One can say
that the fundamental mode on both Bilby and Hawk has the same
phase on the entire surface (and we shall consider by
convention that the wave function is positive, or has a "$0$"
phase, plotted red in Figs.2 and 3). The $2^{nd}$ mode has two
regions of opposite phases and the third has two regions
in-phase separated by a region of opposite phase, etc. The
second and third modes have equal number of nodal lines for
Bilby and Hawk, one nodal line for the second mode and two
nodal lines for the third. Interestingly, the Bilby's forth
mode has three nodal lines, while Hawk's forth mode has only
two nodal lines, as isospectrality does not necessarily imply
an equal number of nodal lines \cite{Gnutzmann}. The two shapes
are isospectral both in the continuous and in the discrete
representations, and eigenfunctions of each shape can be built
from the eigenfunctions of the other by the transplantation
procedure (Appendix A).
 In particular, this allows the extraction of
the phase distribution of the eigenfunctions \cite{Science}.
In the next section, it will be shown that a
correct phase extraction is possible also in the presence of
disorder (up to a certain disorder amplitude).

\section{Disorder effects and robustness of the phase extraction.}

In this section we describe the phase extraction in the
presence of disorder for our isospectral shapes. One can argue
what is more relevant: to consider one single disorder
configuration or the average (of the observables) over a number
of disorder configurations. While both choices have their
relevance, we prefer the second alternative in this paper. Even
in the case of single electron transistors, for instance (or
others transport phenomena where one electron at a time is
involved), a steady value of the current can be read after
thousands or more such single electron events. Temperature
effects or tiny movements of the measurement tip (as in the STM
case) can make each of these electrons to see a slightly
different potential picture. Therefore averaging over a large
number of disorder configurations corresponds to some realistic
experimental conditions. Our plotted results refer to such
averaging over a large number of disorder configurations (1000
for Figs. 6-8). There is also another reason for our choice to
present the results for disorder averaging rather than
individual disorder configurations. The averaging leads to
convergent results, that can be easily reproduced. In fact,
results regarding a single disorder configuration lead to the
same main conclusions and they will be discussed briefly.

\begin{figure}[ht]
\vskip -7.5cm
  \includegraphics[width=12cm]{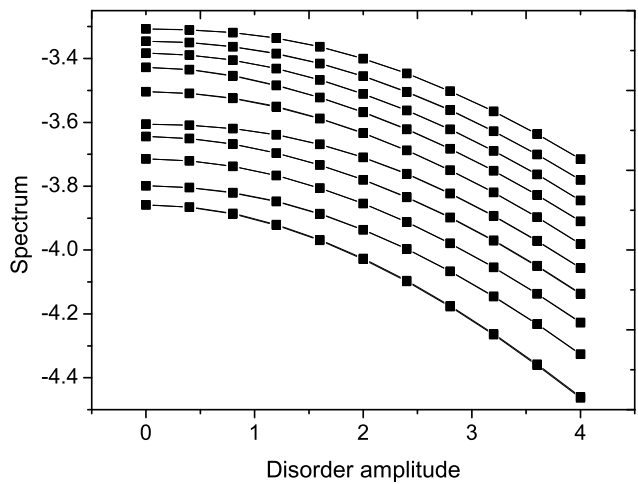}\\
  \centering
  \caption{The Bilby-Hawk eigenvalue spectra (the first 10 eigenvalues) averaged over 1000 disorder
configurations versus increasing disorder amplitude W.
  Notice that the spectra shift from their disorder-free values,
  but the two spectra remain identical to each other (in the limit of numerical errors). }
\end{figure}

Before going to the phase extraction problem -the main focus of
our paper- we present briefly the effect of disorder averaging
on the eigenvalue spectra of Bilby and Hawk in Fig.\,4. When
the Bilby-Hawk spectra are averaged over a large ensemble of
disorder configurations, an expected energy shift is noticed
when the disorder amplitude is increased. The initial spectra
are roughly in the interval [-4:4] according to the theory for
tight-binding model with nearest-neighbors hopping. For the
disorder amplitude W=4 the spectra expands in the interval
roughly $[-4.5:4.5]$ and also the level spacing presents a
monotonic decrease from the bottom of the spectrum towards the
middle (see also \cite{NAZ} and references therein).

It is a known result that the tight-binding spectrum is expanding
with increasing Anderson disorder (see, e.g.  \cite{NAZ} and
references therein). In particular, this means than the lower
eigenvalues move downwards. However, it is important to mention that
the downward evolution of the lower eigenvalues is not just a
mathematical aspect, but should also correspond to the physical
situation: if a surface is affected by random positive and negative
disorder potentials, the lowest eigenfunctions will tend to
supplementary localize around the areas with low potentials,
decreasing the eigenenergies. The effect is not significant for low
disorder amplitudes, that are of interest for the phase extraction.
What is interesting however, is that the Bilby and Hawk spectra
remain very close to each other for any disorder amplitude, as seen
in Fig.4 (the spectral lines for the two shapes practically coincide). The
differences are one order of magnitude smaller than the standard
deviation of the levels statistics. In other words, the
isospectrality is robust against disorder averaging.

The next question, and -as mentioned- the main interest of our
paper, is weather the measured (or "extracted") phase is also robust
against disorder.

To begin with, the wave function square modules were averaged
over a large number of disorder configuration, and we obtained
$\langle|f_{Bi}(i)|^2 \rangle$ and $\langle|f_{Ha}(i)|^2
\rangle$ for Bilby and Hawk. We stress that in the experimental
setup the localization probability is measured, this being the
reason we mediate the square modules. In connection with
experimental procedure \cite{Science} one defines the
"extracted" or "measurable" phase in the following way: it is
the phase distribution that leads to the minimum misfit of the
wave functions after transplantation, as described in the
following numerical algorithm.

\begin{figure}[ht]
\vskip -6.5cm
  \includegraphics[width=11cm]{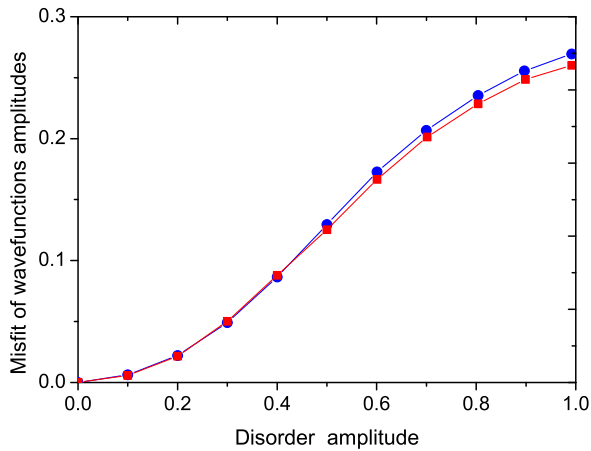}\\
  \centering
  \vskip -0.5cm
  \caption{(Color online) The misfit between the calculated and the "transplantated" eigenfunctions (see description in text)
of Hawk shape, plotted versus the disorder amplitude W.
  The red line with square symbols corresponds to the first eigenvalue, the blue line (with circle symbols) corresponds to the second eigenvalue.}
\end{figure}

We start by assuming the disorder-free phase distribution for
Bilby wave function (the one plotted in Fig.2).  Assuming such initial phase distribution and
the amplitude obtained as the
squared of the disorder average function $\sqrt {\langle|f_{Bi}|^2 \rangle}$ we have now
a complete information about the Bilby wave function and this
is used for transplantation (for transplantation one needs
a full wave function information, both module and phase). The
transplantation procedure produces a Hawk wave function $f_{Ha}^T$
following the recipe given by the Eqs. A5 and A6 in Appendix A.
The square modulus of the transplantated wave function $|f_{Ha}^T(i)|^2$ is compared with
the disorder average Hawk wave function $\langle|f_{Ha}(i)|^2 \rangle$
and we define the misfit as
the differences
between the functions square modules in all sites versus the
sum of the square modules:
\begin{equation}
Misfit=\frac{\sum_i Abs[\langle |f_{Ha}(i)|^2 \rangle - |f_{Ha}^T(i)|^2]}
{\sum_i\big(\langle |f_{Ha}(i)|^2 \rangle + |f_{Ha}^T(i)|^2\big)}
=\frac{\sum_i Abs[\langle |f_{Ha}(i)|^2 \rangle - |f_{Ha}^T(i)|^2]}{2}.
\end{equation}
In the absence of disorder, the misfit should be zero, and the
phase distribution is the one for the unperturbed mode. In the
presence of disorder, the misfit is finite and we have to
search for the phase distribution of the initial Bilby function
that minimize the misfit, repeating the above numerical
algorithm. The convergent solution is the 'extracted' or
measured phase which is obtained when any change of a site
phase would lead to a higher misfit.

The key question is whether this phase coincides with the "unperturbed" phase (which is in fact desired to be measured) or is it
a different phase distribution. The phases that minimize the misfit are determined for
different disorder amplitudes W and are plotted in Figs. 6 and 7 for
the first two modes of Bilby. The corresponding misfit (that
was minimized by these phases) is plotted in Fig.5.

\begin{figure}[ht]
\vskip -9cm
  \includegraphics[width=11cm]{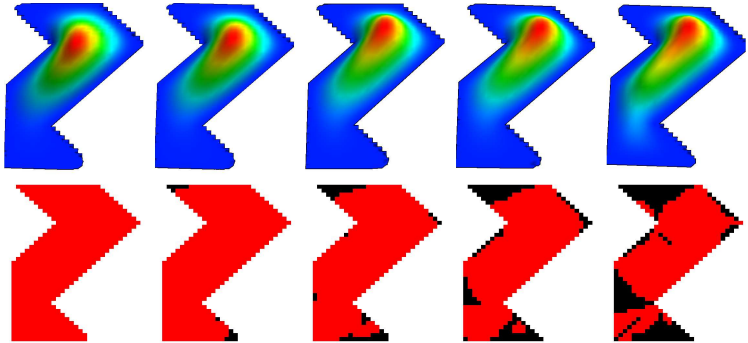}\\
  \centering
  \caption{(Color online) Evolution with increasing disorder of the averaged -over $1000$ disorder configurations- wave function amplitude (first row) and the corresponding "extracted phase" (second row)
  for the first mode of Bilby.
  The disorder amplitude, from left to right, is 0, 0.2, 0.3, 0.4 and 0.7. A significant deviation of the phase from
  the ideal (disorder-free) case can be noticed for disorder higher than $0.3$.}
\end{figure}
\begin{figure}[ht]
\vskip -9cm
  \includegraphics[width=11cm]{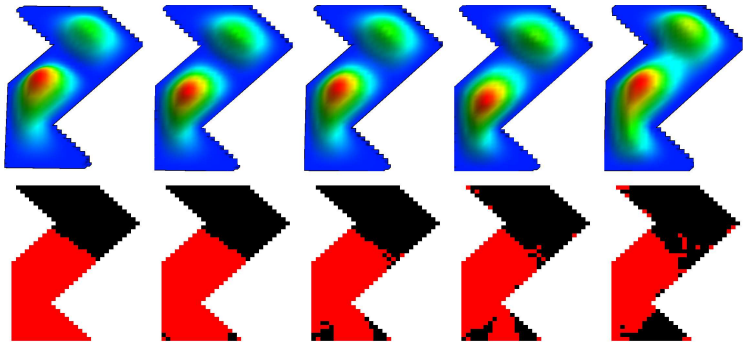}\\
  \centering
  \caption{(Color online) The same as in Fig.6 for the second mode of Bilby.}
\end{figure}

It would be encouraging if even in the presence of the disorder, the best fit
were still realized by the "ideal" phase distribution (the one of the disorder-free system). This
would make its extraction robust.
Here is precisely the point
we want to make in this paper. For low disorder it can be seen
that indeed the "ideal" phase actually ensures the best fit
even if the wave function amplitudes are slightly changed and
a finite misfit appears. Figures 6 and 7 suggest that a
disorder higher than $0.2 - 0.3$ is needed for a significant
variation of the extracted phase from the disorder-free
configuration. The value $0.3$ corresponds to a misfit
of the wave functions of about $5\%$. When the disorder is
increased further, small zones of opposite phase appear inside
the "in-phase" zones (of the unperturbed case) and the nodal
lines shifts. Such an "extracted" phase is no longer related to
the physical phase distribution of the unperturbed mode.

From Figs.\,6 and 7 one can also see that the result of disorder
averaging on the amplitudes of the wave functions is to extend
the zones of high amplitude.

In the captions of Figs. 6 and 7, the disorder amplitude is
expressed in terms of the hopping parameter between
nearest-neighbors, which is taken by convention to be the energy
unit (the usual approach in tight-binding models). However, the
misfit of the wave functions after transplantation, plotted in
Fig.5, is a directly measurable quantity. Therefore Figs.\,6,\,7 and
5 must be combined to give the conclusions in terms of the wave
functions misfit. If the result is expressed in terms of wave
functions misfit, it becomes independent of the energy scale
convention and also independent of the discretization parameter N.

Our numerical results suggest that the
"extracted phase" differs negligibly from the "ideal" phase if the
misfit of the wave functions after transplantation is lower
than a few percents (approx. $4-5\%$, as shown in this section).

We have checked also a large number of single disorder
configurations (and not averages) and one can say that the
general conclusion is the same: the extracted phase differs
significantly from the ideal case when the misfit of the waves
functions (after transplantation) exceeds $5 \%$. In the case
of single disorder configurations one can also talk about the
"intrinsic" phase of a certain eigenmode of the disordered
Bilby (for instance). It is interesting to say that, with
increasing disorder, the intrinsic phase will have shifted
nodal lines, while the extracted phase tends to form new nodal
lines separating small areas of opposite phases, resulting in a
puzzle structure as seen for high disorders in Figs. 6 and 7.

To better connect our numerical simulation with experimental
conditions, let us estimate, for instance (from Fig.4), that a
disorder which shifts the
      energy levels with $10\%$ from the value of the (average) level spacing
      will significantly affect the phase extraction.
      A 2D quantum dot with area of $100nm^2$ may typically have a level spacing
      of $10 meV$. Now if we assume that the thermal oscillations of atoms
      on the surface can be a source of disorder, we can say that
      a shift of the energy levels with $10\%$ from the
       level spacing could be expected for $K_bT=1meV$, resulting approximately
       $T=11K$. As a comparison, the phase extraction in \cite{Science} was carried
       out at $T=4K$.

\vskip 1cm

{\bf Roughness of edges discussion.} In the following we discuss
also another particular form of disorder, namely the roughness of
edges. Let us assume that the edge lines present modification which
consist in including (excluding) adjacent surfaces that do not (do)
belong to the "ideal" shapes Bilby and Hawk, respectively. It is
natural to define the roughness by summing up the total surface
added to the ideal shapes plus the total surface that was eliminated
- and the result should be devised by the total shape surface and
expressed in percents. Fig.8a shows the evolution of the Bilby and
Hawk spectra when mediated over 200 roughness realizations, of
increasing amplitude. The first five eigenvalues of Bilby (plotted
with black) show a slight tendency of moving up, this tendency being
more pronounced for the Hawk eigenvalues (plotted with red). As a
result, the isospectrality is lifted for a roughness exceeding
$2.5\%$.

A comparison between Fig.8a and Fig.4 is not easy to be made, but
some comments are in order.  In Fig.4, the downwards evolution of
the spectrum can be understood in the frame of spectral expansion
under increasing Anderson disorder, the interesting result being the
persistence of isospectrality. On the contrary, the roughness of
edges is not expected to expend the spectrum (the spectrum should
remain in the interval [-4:4], since the diagonal energies were not
modified). For high roughness at edges, isospectrality seems to be
lifted. This should be regarded as a numerical result, which is
plausible if we keep in mind that the edge roughness modifies the
surfaces, and therefor may affect the isospectrality. In terms of
wave functions misfit, isospectrality is lifted for a misfit
exceeding $4\%$, and a roughness exceeding $2.5\%$. Also, from this
value of roughness, the extracted phase shows significant
deviations. For a roughness of $3.5\%$ a large region of "false"
phase can be seen in Fig.8c (the first mode of Bilby was supposed to
be in-phase all over the surface, instead the red area of opposite
phase emerges).

It is important to mention that the results converge remarkably with
those obtained with Anderson disorder, if they are expressed in
terms of wave functions misfit after transplantation. In both cases,
a misfit lower than $4-5\%$ ensures a good phase extraction. In an
experiment, one should always seek to reduce as possible the
disorder, roughness, or other possible errors. However, they can
neither be totally eliminated, nor very accurately estimated. On the
other hand, the misfit of the wave functions (after transplantation)
is inevitably calculated in the phase extraction process, and should
be used as a key indicator. Our numerical simulations suggest that a
misfit lower than $4-5\%$ implies a reliable phase extraction.

\begin{figure}[ht]
\vskip -1cm
  \includegraphics[width=13cm]{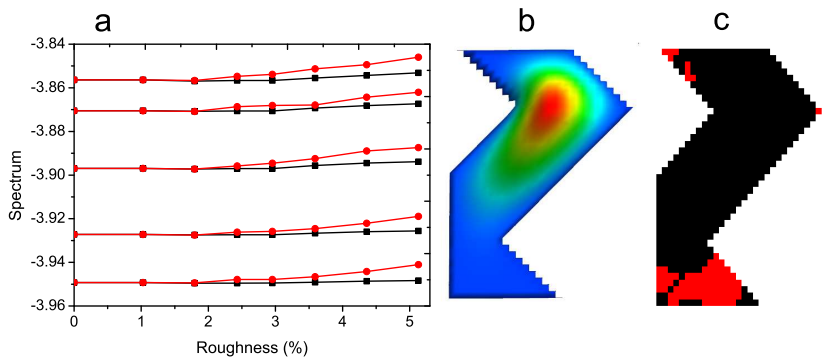}\\
  \centering
  \vskip -10cm
  \caption{(Color online) a) Spectra evolution of  Bilby (black) and Hawk (red) for increasing roughness of edges. b) and c)  amplitude and the extracted phase for the first
  mode of Bilby and a roughness of edges of $3.5 \%$. The spectra and amplitude distributions were calculated by averaging over 200 realizations for each roughness.
  The phase was extracted from the amplitude distributions, with the condition to minimize the misfit after transplantation, as described in text.}
\end{figure}

\section{Conclusions}

When one thinks about phase measurements, interference is the first
word coming to mind - and it was actually the only word for quite a
long time. However, C.R. Moon et.al \cite{Science} -in a remarkable
recent experiment- demonstrated that isospectrality can also be used
to extract phase distributions.

In this paper, we have systematically investigated the robustness of
such a phase extraction. A certain level of disorder or roughness of
edges can compromise the isospectrality-based phase extraction in
the same way in which inelastic scattering or environment-induced
decoherence can compromise the interferometry-based phase
extraction.

Phase extraction in isospectral pairs is possible because the wave
functions of the two shapes can be expressed in term of each other
by the "transplantation" procedure. With disorder, the
transplantation leads to a misfit of the wave functions, which can
be minimized by numerically finding the most suitable phase
distribution - called the "extracted phase". Our numerical results
suggest that, if the misfit is less than $\sim 5\%$, it is likely
that the extracted phase is the correct one, i.e. it coincides with
the phase in the disorder-free case, which is thereby experimentally
available with a certain robustness. If the disorder consists in
roughness of edges, the phase extraction is compromised by a
roughness exceeding $2.5\%$.

The existence of isospectral shapes is a high exception \cite{RMP}
and the experimental realizations of such shapes at the nanoscale
bring supplementary challenges \cite{Science}. In this context, our
proof that the phase extraction can be performed even under
imperfect conditions (quantitative estimations were given) may
hopefully motivate further experimental realizations.

We explicitly present in the paper numerical results
corresponding to averaging over a large number of Anderson disorder configurations or roughness of edges realizarions.

A discrete model is used, which is the natural approach for
disorder analysis, allowing also an easy tailoring of any shape. A proof is provided that isospectrality holds in the
discrete representation, if some general conditions are
fulfilled (see Appendix A).

Another result we obtain is that isospectrality is preserved (in the
limit of statistical fluctuations) if the Bilby and Hawk spectra are
averaged over a large number of Anderson disorder configurations. On
the contrary, if the average is performed over many configurations
of edge roughness, isospectrality is lifted if the roughness exceeds
$2.5\%$.

\section{Acknowledgements}
We acknowledge support from PN-II, Contract TE 90/05.10.2011 and
Core program 45N/2009.

\appendix
\section{Transplantation procedure for the discrete Bilby and Hawk}

In this Appendix, we prove the isospectrality of the discrete Bilby
and Hawk using the transplantation method \cite{Science,RMP}. The
adaptation of the method to the discrete case requires some care due
to the non-locality of the tight-binding Hamiltonian coming from the
hopping terms and, in particular, a transplantation procedure for
the triangles border wave functions will be supplementary needed
(see Eq. A6). Each of the two shapes is divided into 7 triangles
that are labeled as $\alpha$=A,B,...,G for the Bilby drum and as
n=1,2,...,7 for the Hawk drum (see Figs.1,9).

The idea of transplantation is to build a valid eigenfunction of,
say, Hawk, using an eigenfunction of Bilby. This automatically would
prove the one-to-one correspondence of all eigenfunctions (a
bijection), and it will also be shown that they correspond to the
same energy - implying isospectrality. Fig.9 shows schematically how
the transplantation works. An eigenfunction of Hawk is built as
follows (in triangle 1, chosen for exemplification): one adds the
Bilby function from triangles A and F and substracts the function
from triangle E. The triangles in Fig.9 have the borders drawn in
three different colors, the rule being simple: neighboring borders
of two adjacent triangles must have the same color.  The
significance of this rule lies in the continuity conditions at
borders. Then, algebraic summation of the wave functions from
different triangles is performed respecting the "orientation". For
this purpose, the triangles A and F are simply rotated in plane, but
the triangle E must also be flipped once, and as a consequence its
wave function is considered with the sign "-". The full
transplantation recipe, including the rules for the triangles
borders, is given in Eqs. A5 and A6. A legitimate question of the
reader would be why consider this particular recipe ? It is because
it creates indeed valid Hawk eigenfunctions, verifying
$H_{Ha}f_{Ha}=Ef_{Ha}$, as will be proven below. It is beyond our
purpose here to give a general transplantation recipe for any pair
of isospectral shapes, another example can be found in
\cite{Science}, for the Aye-Aye and Beluga shapes, that are devised
in 21 triangles each, etc. It goes without saying that only the
pairs of isospectral shapes allow such transplantation recipes of
building the eigenfunctions of one shape from the other's
eigenfunctions (the existence of a transplantation recipe, that also
does not modify the corresponding eigenenergy automatically implies
isospectrality).

The transplantation relations allow also the phase calculation. In
"perfect" conditions, if one can measure the amplitude of the
eigenmodes inside the triangles $|A|$, $|B|$,...$|G|$ and also
$|1|$, $|2|$,...,$|7|$, then one can use the supplementary relations
given by the transplantation rules
$|A-E+F|=|1|$,......,$|-C+D-F|=|7|$ , to extract uniquely also the
phases of the eigenfunctions in the triangles $A$,$B$, .. ,$G$. In
particular, the eigenfunctions can be chosen real and the phase is
in fact either $0$ or $\pi$, corresponding to positive or negative
sign, respectively (see also the discussion in Section II and in
reference \cite{Science}). In "imperfect" conditions, however, the
equalities $|A-E+F|=|1|$, etc... can only be approximately obeyed,
and what one does is to numerically search for the phase
distributions that minimize the misfit (defined in Eq.2) between the
wave functions calculated by transplantation and those directly
measured. In this paper we simulate the imperfect conditions by
introducing disorder or edge roughness with the purpose to
investigate the robustness of the phase extraction.

\begin{figure}[ht]
\vskip -7cm
  \includegraphics[width=11cm]{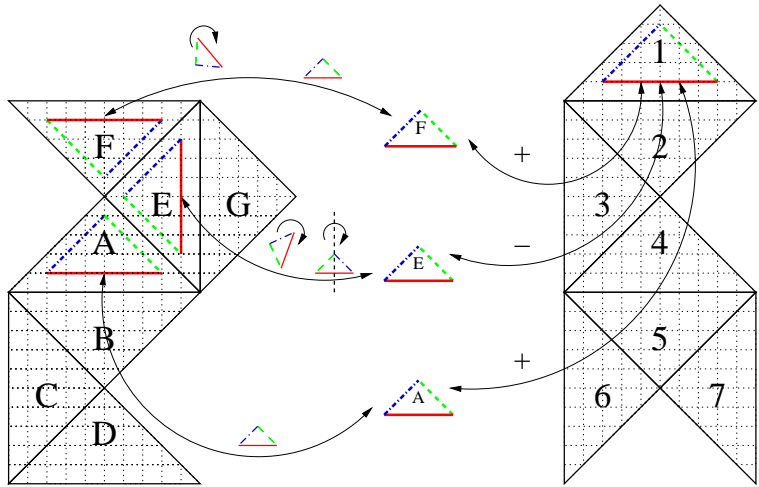}\\
  \centering
  \caption{(Color online) The Hawk eigenfunction in triangle 1 - chosen as example in this plot- ,
  can be expressed as a combination of parts from the corresponding eigenfunction of Bilby, by adding
  the function in triangles A and F and substracting the function in E. }
\end{figure}

In the
discrete model the borders between triangles contain a given number of
sites, and they have to be considered explicitly.
The borders are named with the pair "$\alpha\beta$" meaning the border
between the triangle $\alpha$ and $\beta$ of Bilby, or
"nm" meaning the border between the triangles n and m of Hawk.
First, we write explicitly the two Hamiltonians
writing the parts corresponding to the seven triangles,
to the borders between triangles and the borders-triangles hopping.
The Bilby Hamiltonian can be written:
\begin{equation}
H_{Bi}=\sum_{\alpha=A,..,G}H_\alpha+\frac12\sum_{(\alpha,\beta)}H_{\alpha\beta}+
\sum_{(\alpha,\beta)}(H_{\alpha\rightarrow\alpha\beta}+h.c.),
\end{equation}
Where $H_{\alpha}$ is the triangle Hamiltonian that describes the sites inside
the triangle $\alpha$ and the hopping between them (borders excluded),
$H_{\alpha\beta}$ describes the sites on the border $\alpha\beta$
(the factor $1/2$ ensures that every border is considered
once, $H_{\alpha\beta}=H_{\beta\alpha}$). The last term $H_{\alpha\rightarrow\alpha\beta}$
 describes the hopping between the triangle $\alpha$ and the border $\alpha\beta$
(for $(\alpha,\beta)$  neighboring triangles).
The points on the exterior borders are considered to have infinite potential and need not be included
explicitly in the Hamiltonian.
Consequently, a wave function of Bilby, corresponding
to a certain eigenenergy E, can be written:
\begin{equation}
f_{Bi}=\sum_{\alpha=A,..,G}f_{Bi}^\alpha+\frac12\sum_{(\alpha,\beta)}f_{Bi}^{\alpha\beta},
\end{equation}
and we have:
\begin{equation}
 H_{Bi}~f_{Bi}=E~f_{Bi}.
\end{equation}

In the same way the Hawk Hamiltonian is:
\begin{equation}
H_{Ha}=\sum_{n=1,..,7}H_n+\frac12\sum_{(n,m)}H_{nm}+
\sum_{(n,m)}(H_{n\rightarrow nm}+h.c.).
\end{equation}

Now we start the transplantation procedure meaning that we
build an eigenfunction for Hawk $f_{Ha}$ using a combination of parts from the
Bilby eigenfunction $f_{Bi}$ defined in Eq.A2.
For simplicity we denote by $\alpha$ and $\alpha\beta$ the triangle and border Bilby wave function
$f_{Bi}^\alpha$ and $f_{Bi}^{\alpha\beta}$. Similarly, "n" and "nm" refer to
the triangles and borders projection of the Hawk function $f_{Ha}$.

Now we continue with the full transplantation recipe for our case. The wave functions
in the triangles $n=1,2,...,7$ of Hawk can be built from the Bilby triangle functions as follows (the transplantation matrix is the same as for the
continuous case \cite{Science,RMP}):

\begin{eqnarray}
\left(
  \begin{array}{c}
    1 \\
    2 \\
    3 \\
    4 \\
    5 \\
    6 \\
    7 \\
  \end{array}
\right)=\left(
          \begin{array}{ccccccc}
            1 & 0 & 0 & 0 & -1 & 1 & 0 \\
            0 & 1 & 0 & 0 & 0 & -1 & -1 \\
            0 & 0 & 1 & 0 & -1 & 0 & 1 \\
            -1 & 0 & 0 & 1 & 0 & 0 & -1 \\
            0 & -1 & 0 & -1 & -1 & 0 & 0 \\
            -1 & 1 & -1 & 0 & 0 & 0 & 0 \\
            0 & 0 & -1 & 1 & 0 & -1 & 0 \\
          \end{array}
        \right)\left(
                  \begin{array}{c}
                    A \\
                    B \\
                    C \\
                    D \\
                    E \\
                    F \\
                    G \\
                  \end{array}
                \right),
\end{eqnarray}
where, for instance, $"1"$ refers to the Hawk triangle wave
function $f_{Ha}^1 = P_1f_{Ha}$, $P_1$ being the projection
operator on the triangle "1",  $"A"$ refers to the triangle A
of Bilby, etc.

For the border wave functions we define the following transplantation recipe:

\begin{eqnarray}
\left(
  \begin{array}{c}
    12 \\
    23 \\
    34 \\
    45 \\
    56 \\
    57 \\
  \end{array}
\right)=\left(
          \begin{array}{cccccc}
            1 & 0 & 0 & 0 & 0 & -1 \\
            0 & 1 & 0 & 0 & -1 & 0 \\
            0 & 0 & 1 & -1 & 0 & 0 \\
            -1 & 0 & 0 & 0 & 0 & -1 \\
            0 & 0 & -1 & -1 & 0 & 0 \\
            0 & -1 & 0 & 0 & -1 & 0 \\
          \end{array}
        \right)\left(
                  \begin{array}{c}
                    AB \\
                    BC \\
                    CD \\
                    AE \\
                    EF \\
                    EG \\
                  \end{array}
                \right),
\end{eqnarray}
where $"12"$ refers to the triangle wave function $f_{Ha}^{12} =
P_{12}f_{Ha}$, $P_{12}$ being the projection operator on the border
"12". The above relation results
from the general rules (A5) considering that the wave functions
vanish on the external borders. The transplantation matrix for the border wave function
has the dimension $6$ equal to the number of internal
triangle borders.

When performing the operations (A5) and (A6), it is important
to respect the "paper folding" principle, exactly as in the
continuous case \cite{Berard,Chapman,RMP}. For the first (A5)
equation, this means that one has to (imaginary) fold the triangle E over
the triangle A on the common border and extract in each point
from the triangle A wave function the corresponding triangle E
function that landed upon it after the folding, etc.

In the following we have to prove that the Hawk wave function $f_{Ha}$ as described in Eq. (A5) and (A6)
is a proper wave function of the Hamiltonian $H_{Ha}$ , and corresponding to the same energy E as in Eq.(A3).
For this purpose, we apply
the Hawk Hamiltonian (A4) to the transplantated wave function. The proof is a bit lengthly
to be written in totality, but rather straightforward.
Let us consider for instance the projection on triangle labeled "1" of $H_{Ha}f_{Ha}$:
\begin{eqnarray}
P_1H_{Ha}f_{Ha}&=&H_{1}f_{Ha}^{1}+H_{12\rightarrow 1}f_{Ha}^{12}\\
 &=&H_{1}(f_{Bi}^{A}-f_{Bi}^{E}+f_{Bi}^{F})+H_{12\rightarrow
1}(f_{Bi}^{AB}-f_{Bi}^{EG}).\nonumber
\end{eqnarray}

On the other hand on the Bilby shape we have:

\begin{eqnarray}
&a)&H_{A}f_{Bi}^{A}+H_{AE\rightarrow A}f_{Bi}^{AE}+H_{AB\rightarrow A}f_{Bi}^{AB}=Ef_{Bi}^{A},\nonumber\\
&b)&H_{E}f_{Bi}^{E}+H_{EG\rightarrow E}f_{Bi}^{EG}+H_{EF\rightarrow E}f_{Bi}^{EF}
+H_{AE\rightarrow E}f_{Bi}^{AE}=Ef_{Bi}^{E},\nonumber\\
&c)&H_{F}f_{Bi}^{F}+H_{EF\rightarrow
F}f_{Bi}^{EF}=Ef_{Bi}^{F}.
\end{eqnarray}

Now we have to extract Eq.A8(b) from Eq.A8(a) and add Eq.A8(c).
 The necessary ingredient for isospectrality is that all triangle Hamiltonians are identical:
$H_{1}\equiv H_{2}\equiv ... H_{7}\equiv H_{A}\equiv ... \equiv H_{G}$,
and also those corresponding to similar borders and borders-triangles hopping: $H_{12}\equiv H_{AB}$, $H_{12\rightarrow 1}\equiv H_{AB\rightarrow A}$, etc.

The desired result is obtained:
\begin{equation}
P_1H_{Ha}f_{Ha}=Ef_{Ha}^{1}.
\end{equation}
For the other triangles and for the borders the proof runs
identically, therefore one can write:
\begin{equation}
H_{Ha}f_{Ha}=Ef_{Ha}.
\end{equation}

One should notice that the isospectrality proof given here is rather general and does not depend on
the particular choice of the discrete lattice (as the discrete square network used in our numerical calculation).
The only condition is
the mentioned equivalence of sub-systems Hamiltonians.

\section {Comparison between discrete and continuous models}

As shown in the Appendix A, the discrete Bilby and Hawk are
isospectral for any discrete representation (see the general
conditions in Appendix A). So there is no need to approach the
continuous limit (by increasing the number of sites) in order
to achieve isospectrality. Still, it is instructive to show
that the continuous limit is approached relatively easy, if one
aims to study the first few energy levels (there is no need for
very many sites or a high computing power). One possible
criterium for approaching the continuous limit satisfactory can
be a similar spectral structure for the first few energy levels
of interest. In other words, the ratio between level spacings
should be very close to the corresponding ratio for the
continuous model.

In Fig.10 we plot the first six level spacings divided by the first
(energy distance between eigenvalues $1^{st}$ and $2^{nd}$). One can
see that, for N=13 the ratios are already very close for discrete
and continuous models, while for N=21, the difference is even less,
as expected. N=13 corresponds to 217 sites on each shape, and N=21
to 641 sites \cite{sites}. The continuous spectrum was taken from
\cite{Wu,Driscoll}.

\begin{figure}[ht]
\vskip -4cm
  \includegraphics[width=9cm]{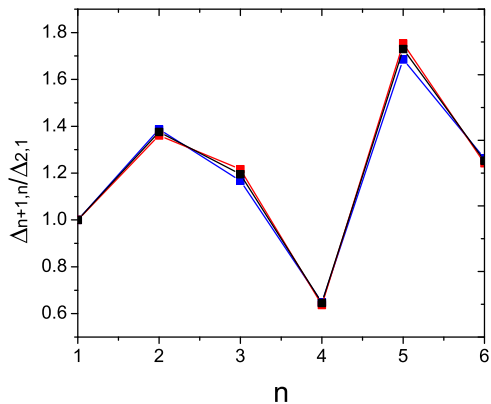}\\
  \centering
  \caption{(Color online) The ratio between level spacings $\Delta E_{n,n+1}/\Delta E_{1,2}$ (of the Bilby-Hawk isospectral shapes) is plotted versus
  the eigenvalue index $n$. The red  (light gray) line corresponds to the continuous model,
  the blue (dark gray) and black lines correspond to the discrete model with $N=13$ and $N=21$, respectively.}
\end{figure}

Some more comments can be made regarding the comparison between
discrete and continuous models. In \cite{Marian}, it is shown
that the discrete and continuous models give the very same
result also for dynamical quantities, such as the
time-dependent magnetization. The technical difference between
continuous and discrete models consist in the the
approximations made: in the continuous model a finite number of
(analytically known) triangle modes are used to compute the
Bilby-Hawk spectrum, while in our discrete approach a finite
number of sites is used for the discretization of space. The
discrete model is more suitable for implementation of disorder
(as we do in this paper) or -eventually- the electron-electron
interaction, as we plan to do in a future work. Also, the
discrete model allows for easy tailoring of any shapes, without
the need of analytical knowledge of sub-systems spectra. The
discrete model is also particularly suitable for the simulation
of the experimental phase extraction procedure as in Moon et.
al \cite{Science}.


\begin{thebibliography}{99}
\bibitem{comment1}The
interference method can be regarded as measuring the relevant phases
of the scattering eigenmodes, while the isospectrality method determines the phase
distribution of the eigenmodes on a surface. In this respect, the
difference is basically between the measured systems.
\bibitem{sites}The total number of sites (for both Bilby and Hawk), corresponding to $N$ sites
 on the hypothenusis of an elementary triangle is: $\frac74(N-3)^2+2(2N-5)$.
 Notice that, in our model, an odd value is assumed for $N$. For instance in Fig.1 we have $N=7$ corresponding to $46$ internal points
 (the points on the external borders are not counted, as they have infinite potential).
\end{thebibliography}
\end{document}